\documentclass{article}

\usepackage[dblblindworkshop,final]{neurips_2025}

\workshoptitle{MATH-AI}

\usepackage[utf8]{inputenc} % allow utf-8 input
\usepackage[T1]{fontenc}    % use 8-bit T1 fonts
\usepackage{hyperref}       % hyperlinks
\usepackage{url}            % simple URL typesetting
\usepackage{booktabs}       % professional-quality tables
\usepackage{amsfonts}       % blackboard math symbols
\usepackage{nicefrac}       % compact symbols for 1/2, etc.
\usepackage{microtype}      % microtypography
\usepackage{xcolor}         % colors
\usepackage{multicol}

\usepackage{listings}

\definecolor{keywordcolor}{rgb}{0.7, 0.1, 0.1}   % red
\definecolor{tacticcolor}{rgb}{0.0, 0.1, 0.6}    % blue
\definecolor{commentcolor}{rgb}{0.3, 0.5, 0.3}   % grey
\definecolor{symbolcolor}{rgb}{0.0, 0.1, 0.6}    % blue
\definecolor{sortcolor}{rgb}{0.1, 0.5, 0.1}      % green
\definecolor{attributecolor}{rgb}{0.7, 0.1, 0.1} % red
\definecolor{rulecolor}{rgb}{0, 0, 0}

\usepackage{graphicx}

\usepackage{placeins}

\title{Kimina Lean Server: A High-Performance Lean Server for Large-Scale Verification}
\workshoptitle{Kimina Lean Server: A High-Performance Lean Server for Large-Scale Verification}

\author{
  Marco Dos Santos$^{\,1,2}$ \\
  \And
  Hugues de Saxc\'e$^{\,1}$\\
  \And
  Haiming Wang$^{\,3}$ \\
  \And
  Ran Wang$^{\,1}$ \\
  \AND
  Mantas Bak\v{s}ys$^{\,1,2}$ \\
  \And
  Mert \"{U}nsal$^{\,1}$ \\
  \And 
  Junqi Liu$^{\,3}$ \\
  \And
  Zhengying Liu$^{\,3}$ \\
  \And 
  Jia Li$^{\,1}$ \\
  \AND
  \textnormal{$^{1\,}$Project Numina}
  \hspace{5pt}
  \textnormal{$^{2\,}$University of Cambridge}
  \hspace{5pt}
  \textnormal{$^{3\,}$Moonshot AI}
}

\begin{document}

\maketitle

\begin{abstract}
Recent progress in neural theorem proving has been driven by the training of large language models on Lean 4 problems via reinforcement learning, a process that requires fast and scalable verification of proofs. 

We introduce the Kimina Lean Server, an open-source project designed as a high-performance verifier for reinforcement learning pipelines. Built on top of the Lean REPL (Read-Eval-Print Loop) maintained by the Lean FRO\footnote{The Lean Focused Research Organization (FRO) is a non-profit organization to improve Lean's critical systems and guide it toward long-term self-sustainability, cf. \href{https://lean-lang.org/fro/about}{About - The Lean FRO}}, our server combines server-side parallelism by managing multiple Lean processes in parallel with a Least Recently Used (LRU) caching mechanism that reuses Lean imports across requests. On the client side, a lightweight Python package enables submitting proof batches and receiving Lean feedback, including extracted tactics and tactic states.

Together, these features enable a scalable workflow for large-scale verification and data extraction. In our experiments, the Kimina Lean Server outperforms previous Lean interaction tools, achieving a 1.5 to 2 times speedup in verification time. Moreover, its improved efficiency has enabled its use in the large-scale training of state-of-the-art models such as Kimina-Prover \citep{kimina-prover}.

We hope that our open-source project\footnote{Our code is available at \url{https://github.com/project-numina/kimina-lean-server}.} will support the neural theorem proving community and accelerate future progress by enabling efficient large-scale verification and proof data extraction.
\end{abstract}

\section{Introduction}
Recent advances in AI for formal mathematics have driven the development of systems that integrate machine learning models with interactive proof assistants. These interactive environments enable using proof outcomes as reward signals for reinforcement learning.  With its active community and robust ecosystem, Lean 4 \citep{lean4paper} has become the proof assistant of choice for both mathematicians and AI researchers. The successful large-scale reinforcement learning training of theorem proving models \citep{kimina-prover, deepseek-prover-v2, seed-prover} highlights a  community need for a highly efficient Lean server which supports large-scale verification and data extraction workflows.

Several projects have tackled the challenge of interfacing Lean 4 with Python \citep{yang2023leandojo, pantograph2025, leanclient2025, leaninteract, thakur2025proofwala}, each with distinct trade-offs. While a lot of progress has been made in the past couple of years, existing tools face performance bottlenecks and scalability issues. For instance, most are not designed to support parallelization across CPU cores and incur significant initialization costs for each verification. While tools such as leanclient \citep{leanclient2025} and LeanInteract \citep{leaninteract} support parallelization, their performance is limited, and a high-performance solution designed specifically for large-scale reinforcement learning pipelines has been lacking.

The Kimina Lean Server is designed to address this gap. It performs server-side parallelization of verification and extraction tasks, and it employs a Least Recently Used caching strategy that reuses Lean imports across multiple requests, reducing initialization costs and enabling large-scale batch processing. On the client side, we provide a lightweight Python package that makes it easy to submit multiple Lean scripts and receive Lean feedback programmatically. In addition, our extraction pipeline processes Lean's infotree output to partition proofs into non-overlapping tactics, which is particularly useful for tree search models.

Our contributions include:
\begin{itemize}
\item \textbf{Server-side parallelization.} Parallel Lean 4 verification across multiple Lean REPL processes to fully utilize multicore CPUs, enabling efficient batch processing of Lean code.
\item \textbf{LRU caching of imported modules.} In-memory caching of frequently used modules (e.g., Mathlib) to considerably reduce initialization overhead.
\item \textbf{Lightweight client-side Python package.} A simple, high-level API to submit Lean scripts and receive Lean feedback, designed for integration with RL and data-extraction pipelines. 
\item \textbf{Data extraction for tree search.} Our client package can be used to process Lean's infotree output to provide partitions of proofs with non-overlapping tactics and tactic states, a format tailored for tree search models.
\end{itemize}

\begin{table}
  \caption{Verification time (mm:ss) for the 9,419 Lean proofs from NuminaMath-LEAN (lower is better). Our Kimina Lean Server is fastest across all core counts, outperforming the next best baseline by a factor of 1.5 to 2.}
  \label{tab:performance}
  \centering
  \begin{tabular}{lccccc}
    \toprule
    Project & 8 cores & 16 cores & 32 cores & 64 cores \\
    \midrule
    leanclient & 109:55 & 56:58 & 30:16 & 18:01 \\
    LeanInteract & 87:35 & 45:51 & 24:11 & 12:56 \\
    Kimina Lean Server & \textbf{42:40} & \textbf{21:48} & \textbf{11:33} & \textbf{7:56} \\
    \bottomrule
  \end{tabular}
\end{table}

\section{Lean Server}
\subsection{Server Side}
The server exposes a Representational State Transfer (REST) API that enables proof verification from any programming language. To achieve high performance, its architecture is built on two core principles: server-side parallelization and import caching.

\paragraph{Parallelization}
The server maintains a pool of Lean REPLs, each running in its own process to ensure optimal CPU core usage. As requests arrive, the server routes each proof to an idle Lean REPL and returns the response. This parallelization strategy enables the server's performance to scale efficiently with available hardware, as demonstrated in our experiments in Section~\ref{section:experiments}.

\begin{figure}[ht]
  \centering
  \includegraphics[width=\linewidth]{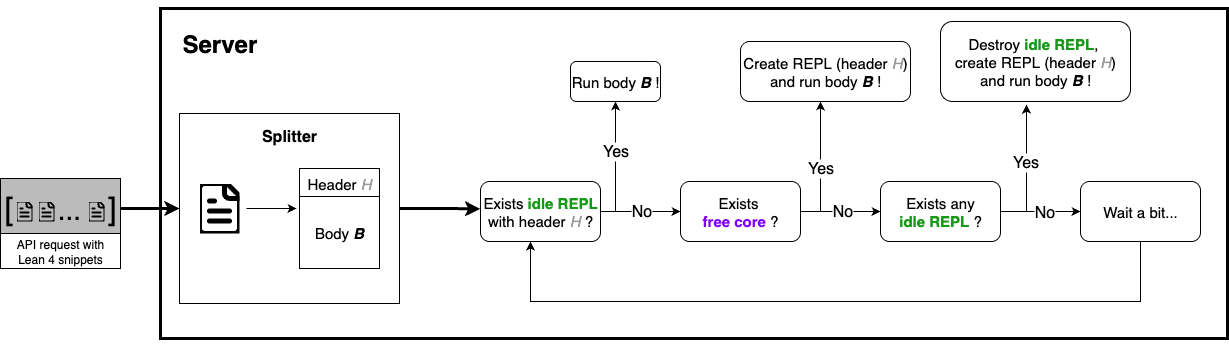}
  \caption{Architecture of the Kimina Lean Server parallelization and caching.}
  \label{fig:parallel-arch}
\end{figure}

\paragraph{Caching}
Initializing a Lean REPL is computationally expensive, primarily due to the time required to load large libraries such as \texttt{Mathlib}. To mitigate this overhead, we introduce a caching mechanism that reuses REPLs with pre-loaded imports.

As shown in Figure~\ref{fig:parallel-arch}, each incoming script is split into a header \textbf{\textit{H}} containing only imports and a body \textbf{B} with the remaining code. The server then uses the header as a key in a Least Recently Used (LRU) cache to find a "warmed" worker that has already processed the same imports. If a matching worker is found, the server only needs to verify the body of the script, reusing the ready context. This mechanism leads to a significant performance boost, as quantified in our experiments in Section~\ref{section:experiments}.

\subsection{Client Side}
\paragraph{Client-Server Interaction}
To simplify interaction from Python, we provide a lightweight client as a PyPI package that wraps the server's standard REST API. All interactions are handled by a single call, \texttt{check}, which takes a list of Lean scripts and returns a list of results. The server handles all parallel execution and REPL caching behind the scenes.

For each script, the response includes any Lean messages (warnings or errors), the REPL environment identifier, the elapsed time, and an optional infotree. This infotree can be used to extract tactics and tactic states, as described in the next paragraph.

An example of large-scale verification is given in Appendix~\ref{appendix:verification}.

\paragraph{Data Extraction via Infotree Processing}
To support data extraction and interactive proof search, our Python client package can process Lean's infotree to extract a clean sequence of non-overlapping tactics and their corresponding tactic states.

Our processing pipeline first extracts the position of each tactic, adjusts these positions to eliminate overlaps, extracts the corresponding code snippet for each interval, and performs final adjustments to handle whitespaces, comments, and special tactics. This produces a clean sequence of tactics and states, supporting all Lean tactics, including \texttt{have} and \texttt{let} tactics, \texttt{calc} mode, and \texttt{conv} mode, which are not always supported by the LeanREPL's Tactic mode.

An example of proof data extraction is given in Appendix~\ref{appendix:data-extraction}.

\section{Experiments}
\label{section:experiments}
To validate the design and performance of our Kimina Lean Server, we conducted a series of experiments focusing on three key aspects: overall performance compared to existing tools, scalability with increasing CPU cores, and the efficiency gains from our caching mechanism. Finally, we report on its successful application in a large-scale reinforcement learning workflow.

For all benchmarks, experiments were conducted on a Google Cloud Platform (GCP) C4 general-purpose virtual machine, powered by 5th Generation Intel Xeon processors, using Lean v4.15.0. Further details on our experimental setup are provided in Appendix~\ref{appendix:experiments}.

\paragraph{Performance}
We first evaluated our server's performance against two other available tools that support some level of parallelization: \texttt{leanclient} \citep{leanclient2025} and \texttt{LeanInteract} \citep{leaninteract}. We measured the total time required to process a large dataset of proofs, scaling the number of available CPU cores. For this benchmark, we used a filtered subset of the NuminaMath-LEAN dataset \citep{kimina-prover}, which contains 9,419 valid, sorry-free proofs.

The results, summarized in Table~\ref{tab:performance}, show that our server consistently outperforms both baselines across all core counts. On average, our server is 1.5 to 2 times faster than the next best alternative, demonstrating its superior efficiency for large-batch proof verification.

\paragraph{Parallelization and Scalability}
A key design goal of our server is to maximize hardware utilization. To demonstrate its scalability, we benchmarked its performance on the full NuminaMath-LEAN dataset while varying the number of available CPU cores from 8 to 64. A single Lean REPL process is single-threaded, and its CPU usage does not exceed one core. Our server leverages this by running one persistent REPL process per dedicated core, enabling highly efficient parallelization.

The results in Table~\ref{tab:cpu-scaling} show a strong correlation between added cores and reduced verification time. Scaling from 8 to 32 cores reduces the total time from 42:40 to 11:33, a nearly 4x speedup. This demonstrates our server's ability to effectively scale with available computational resources.

\begin{table}[h]
  \caption{Performance scaling of proof verification with different CPU configurations on the NuminaMath-LEAN dataset \citep{kimina-prover}.}
  \label{tab:cpu-scaling}
  \centering
  \begin{tabular}{ccc}
    \toprule
    \textbf{\# CPUs} & \textbf{Total Verification Time (mm:ss)} & \textbf{Average Verification Time (s)} \\
    \midrule
    8  & 42:40 & 0.272 \\
    16 & 21:48 & 0.139 \\
    32 & 11:33 & 0.074 \\
    64 & 7:56 & 0.051 \\
    \bottomrule
  \end{tabular}
\end{table}

\paragraph{Caching}
To quantify the impact of our LRU caching mechanism, we compared two modes on the NuminaMath-LEAN dataset: a "non-cached" mode (fresh REPL per proof) and a "cached" mode (reusing pre-initialized REPLs).

The results in Table~\ref{tab:caching-performance} show that caching reduces the average verification time from 0.099 seconds to 0.051 seconds, a 1.94x speedup. This demonstrates the important performance benefit of avoiding the costly re-importing of large libraries, an advantage particularly effective for datasets and workflows that frequently reuse the same imports.

\begin{table}[h]
  \caption{Performance comparison of cached vs. non-cached verification on the 9,419 Lean proofs from NuminaMath-LEAN with 64 CPU cores. Caching leads to significantly faster verification times.}
  \label{tab:caching-performance}
  \centering
  \begin{tabular}{ccc}
        \toprule
        \textbf{Mode} & \textbf{Total Verification Time (mm:ss)} & \textbf{Average Verification Time (s)} \\
        \midrule
        Cached     & 7:56 & 0.051 \\
        Non-Cached & 15:28 & 0.099 \\
        \bottomrule
    \end{tabular}
\end{table}

\paragraph{Validation in Large-Scale Training}
Beyond synthetic benchmarks, our Lean server has been tested as the core verification engine for the reinforcement learning training of the Kimina-Prover family of models \citep{kimina-prover}. The successful training of these models, which have twice achieved state-of-the-art performance on the miniF2F benchmark \citep{minif2f}, serves as a practical validation of our system's efficiency and robustness for neural theorem proving workflows.

\section{Conclusion}
We introduced the Kimina Lean Server, an open-source Lean server designed for high-performance proof verification in large-scale machine learning workflows. Our server combines server-side parallelism to fully utilize multi-core systems with an LRU caching mechanism that reuses costly module imports across requests.

To facilitate interaction, we provide a lightweight Python package. A single \texttt{check} function enables users to submit batches of proofs and receive structured feedback from Lean, abstracting away the complexity of the underlying parallel execution and caching. Our server outperforms existing parallel solutions by 1.5 to 2 times, and has served as the core verification engine for training state-of-the-art theorem-proving models, further validating its efficiency and scalability. 

With its direct use of the official Lean REPL, our server is compatible with any Lean version supported by the REPL. Furthermore, the architecture is modular: with minor modifications to the process spawning logic, the server could be adapted to use alternative Lean proof checkers while retaining the same REST API.

We hope that our open-source server will benefit the AI for mathematics community by drastically decreasing the effort required to interact with Lean from Python, thereby accelerating future research.

\bibliography{main}

@misc{leanrepl,
  author = {{Lean FRO}},
  title = {{A read-eval-print-loop for Lean 4}},
  year = {2023},
  publisher = {GitHub},
  journal = {GitHub repository},
  howpublished = {\url{https://github.com/leanprover-community/repl}}
}

@InProceedings{pantograph2025,
author={Aniva, Leni
and Sun, Chuyue
and Miranda, Brando
and Barrett, Clark
and Koyejo, Sanmi},
editor={Gurfinkel, Arie
and Heule, Marijn},
title={{Pantograph: A Machine-to-Machine Interaction Interface for Advanced Theorem Proving, High Level Reasoning, and Data Extraction in Lean 4}},
booktitle={Tools and Algorithms for the Construction and Analysis of Systems},
year={2025},
publisher={Springer Nature Switzerland},
address={Cham},
pages={104--123},
isbn={978-3-031-90643-5},
doi= {10.1007/978-3-031-90643-5_6}
}

@inproceedings{yang2023leandojo,
    title={{{LeanDojo}: Theorem Proving with Retrieval-Augmented Language Models}},
    author={Yang, Kaiyu and Swope, Aidan and Gu, Alex and Chalamala, Rahul and Song, Peiyang and Yu, Shixing and Godil, Saad and Prenger, Ryan and Anandkumar, Anima},
    booktitle={Neural Information Processing Systems (NeurIPS)},
    year={2023}
}

@misc{thakur2025proofwala,
      title={${\rm P{\small ROOF}W{\small ALA}}$: Multilingual Proof Data Synthesis and Theorem-Proving},
      author={Amitayush Thakur and George Tsoukalas and Greg Durrett and Swarat Chaudhuri},
      year={2025},
      eprint={2502.04671},
      archivePrefix={arXiv},
      primaryClass={cs.AI},
      url={https://arxiv.org/abs/2502.04671},
}

@software{leanclient2025,
  author = {Oliver Dressler},
  title = {{leanclient: Python client to interact with the lean4 language server}},
  url = {https://github.com/oOo0oOo/leanclient},
  month = {1},
  year = {2025}
}

@misc{mathlib-port-status,
  title = {{Mathlib Port Status}},
  author = {{Lean Community}},
  year = {2023},
  url = {https://leanprover-community.github.io/mathlib-port-status/}
}

@inproceedings{lean4paper,
author = {Moura, Leonardo de and Ullrich, Sebastian},
title = {{The Lean 4 Theorem Prover and Programming Language}},
year = {2021},
isbn = {978-3-030-79875-8},
publisher = {Springer-Verlag},
address = {Berlin, Heidelberg},
url = {https://doi.org/10.1007/978-3-030-79876-5_37},
doi = {10.1007/978-3-030-79876-5_37},
booktitle = {Automated Deduction - CADE 28: 28th International Conference on Automated Deduction, Virtual Event, July 12-15, 2021, Proceedings},
pages = {625-635},
numpages = {11}
}

@article{kimina-prover,
	title = {{Kimina-Prover Preview: Towards Large Formal Reasoning Models with Reinforcement Learning}},
	author = {Wang, Haiming and Unsal, Mert and Lin, Xiaohan and Baksys, Mantas and Liu, Junqi and {Dos Santos}, Marco and Sung, Flood and Vinyes, Marina and Ying, Zhenzhe and Zhu, Zekai and Lu, Jianqiao and {de Saxc{\'e}}, Hugues and Bailey, Bolton and Song, Chendong and Xiao, Chenjun and Zhang, Dehao and Zhang, Ebony and Pu, Frederick and Zhu, Han and Liu, Jiawei and Bayer, Jonas and Michel, Julien and Yu, Longhui and Dreyfus-Schmidt, L{\'e}o and Tunstall, Lewis and Pagani, Luigi and Machado, Moreira and Bourigault, Pauline and Wang, Ran and Polu, Stanislas and Barroyer, Thibaut and Li, Wen-Ding and Niu, Yazhe and Fleureau, Yann and Hu, Yangyang and Yu, Zhouliang and Wang, Zihan and Yang, Zhilin and Liu, Zhengying and Li, Jia},
	year = {2025},
	url = {http://arxiv.org/abs/2504.11354},
}

@software{leaninteract,
  author = {Poiroux, Auguste and Kuncak, Viktor and Bosselut, Antoine},
  title = {{LeanInteract: A Python Interface for Lean 4}},
  url = {https://github.com/augustepoiroux/LeanInteract},
  year = {2025}
}

@article{deepseek-prover-v2,
  title={Deepseek-prover-v2: Advancing formal mathematical reasoning via reinforcement learning for subgoal decomposition},
  author={Ren, ZZ and Shao, Zhihong and Song, Junxiao and Xin, Huajian and Wang, Haocheng and Zhao, Wanjia and Zhang, Liyue and Fu, Zhe and Zhu, Qihao and Yang, Dejian and others},
  journal={arXiv preprint arXiv:2504.21801},
  year={2025}
}

@article{seed-prover,
  title={{Seed-prover: Deep and broad reasoning for automated theorem proving}},
  author={Chen, Luoxin and Gu, Jinming and Huang, Liankai and Huang, Wenhao and Jiang, Zhicheng and Jie, Allan and Jin, Xiaoran and Jin, Xing and Li, Chenggang and Ma, Kaijing and others},
  journal={arXiv preprint arXiv:2507.23726},
  year={2025}
}

@inproceedings{
minif2f,
title={{miniF2F: a cross-system benchmark for formal Olympiad-level mathematics}},
author={Kunhao Zheng and Jesse Michael Han and Stanislas Polu},
booktitle={International Conference on Learning Representations},
year={2022},
url={https://openreview.net/forum?id=9ZPegFuFTFv}
}

\newpage

\appendix

\section{Related Work}
Over the past couple of years, a number of projects have tackled the challenge of interfacing Lean 4 with Python, each addressing different needs and exhibiting distinct trade-offs. Lean 4 was released in 2021 \citep{lean4paper} and the community has since migrated the entirety of Mathlib to this new version \citep{mathlib-port-status}. The Lean community has also developed a number of tools to facilitate the interaction with Lean 4 from Python. These tools are designed to support various tasks, including data extraction and interaction with Lean's proving environment.

The primary low-level interface for programmatic interaction is the official LeanREPL \citep{leanrepl}, a read-eval-print loop that exposes Lean as an interactive subprocess. This interface allows submitting one tactic at a time and obtaining the new goals, or sending an entire proof script and receiving the final result. LeanREPL does not support multi-line \texttt{have} and \texttt{let} tactics, \texttt{calc} mode, or \texttt{conv} mode natively. Pantograph \citep{pantograph2025} was developed as a more feature-rich alternative, overcoming some of LeanREPL's limitations by treating subgoals independently and supporting \texttt{have}, \texttt{let}, \texttt{conv}, and \texttt{calc} modes, as well as extracting tactic states and whole proof scripts with comments. Both of these tools are inherently single-threaded and are not designed for the high-throughput verification required by large-scale reinforcement learning pipelines.

To address the need for large-scale verification, several tools have focused on enabling parallel processing of Lean code. leanclient \citep{leanclient2025} interfaces with Lean via its Language Server Protocol (LSP) to support batch processing of files in parallel. LeanInteract \citep{leaninteract} offers a similar capability by managing multiple REPL instances. Although these tools support parallel processing, their performance remains limited for large-scale training, and they are not designed to extract proof data such as individual tactics and tactic states.

Other projects like ProofWala \citep{thakur2025proofwala} and LeanDojo \citep{yang2023leandojo} provide more comprehensive gym-like environments for data extraction and proof search. These tools are not optimized for the verification speed needed in modern RL workflows.

Our work builds on these previous efforts by providing a solution specifically engineered for high-performance verification and data extraction, addressing the scalability needs of training state-of-the-art language models.

\section{Experiments}
\label{appendix:experiments}
\subsection{Experimental Setup}
All experiments were conducted on a Google Cloud Platform (GCP) C4 general-purpose virtual machine. We selected the C4 instance type, powered by 5th Generation Intel Xeon processors, as it is optimized for CPU-based inference.

The machine was configured with 72 vCPUs, 540 GB of RAM (7.5 GB per vCPU), and a 200 GB Hyperdisk Balanced storage volume with 20,000 provisioned IOPS (Input/Output operations Per Second). To ensure predictable performance for these compute-bound tasks, we disabled Simultaneous Multi-Threading (SMT), which can hinder overall application performance and add unpredictable variance to jobs. 

All experiments were run fully locally to eliminate any network latency.

The software environment consisted of a standard Linux distribution with Miniconda, Git, elan, and Lean v4.15.0, which was used for all benchmarks.

\subsection{Dataset Processing}
Our benchmark dataset is a filtered subset of the NuminaMath-LEAN dataset \citep{kimina-prover}. We performed a three-step processing pipeline to ensure data quality and validity:

\begin{enumerate}
    \item First, we processed the original dataset to remove duplicate entries based on their unique identifiers (column \texttt{uuid}).
    \item Next, we filtered for complete proofs (column \texttt{ground\_truth\_type} equal to \texttt{complete}).
    \item Finally, we performed a validation step and removed a small number of proofs that either contained \texttt{sorry} (26 proofs) or failed to compile with our target Lean version (3 proofs).
\end{enumerate}

This process resulted in a final dataset of 9,419 valid, sorry-free proofs.

\subsection{Step-by-Step Protocol}
To ensure a fair comparison, we followed a specific protocol for our server and the two baselines, \texttt{leanclient} and \texttt{LeanInteract}. The Python scripts used to run the \texttt{leanclient} and \texttt{LeanInteract} experiments are available in the supplementary material.

\paragraph{Kimina Lean Server}
The server was configured by setting environment variables to match the core count of each experiment (e.g., \texttt{LEAN\_SERVER\_MAX\_REPLS}=8, 16, 32, 64). We also increased the request timeout (\texttt{LEAN\_SERVER\_MAX\_WAIT}) and memory-per-REPL (\texttt{LEAN\_SERVER\_MAX\_REPL\_MEM}). The benchmark was executed using the provided Python client, which submits all 9,419 proofs in batches to the server API.

\paragraph{leanclient}
For \texttt{leanclient}, we created a new Lake project and configured its \texttt{lean-toolchain} and \texttt{lakefile.lean} to use the exact same Lean and Mathlib versions (v4.15.0) as our server. The verification was then performed using a Python script that interacted with this Lake project.

\paragraph{LeanInteract}
For \texttt{LeanInteract}, we used a script that followed the library's official recommendation for parallel processing: using one \texttt{LeanREPLConfig} instance, and one \texttt{AutoLeanServer} instance per process. We did not specify \texttt{add\_to\_session\_cache=True} when calling \texttt{run} on a command because each proof gets its own \texttt{AutoLeanServer}, and having to manage pools of reusable \texttt{AutoLeanServer} instances would involve exactly the engineering effort put into our Lean server. 

\section{Example Usage}
We illustrate the two main use cases of our Kimina Lean Server: large-scale verification of Lean scripts and structured extraction of proof data.

\subsection{Large-scale Verification}
\label{appendix:verification}
This example demonstrates how to use the Kimina Lean Server to verify a large batch of Lean scripts. The simplest method is to use the high-level \texttt{run\_benchmark} function, which handles data loading and processing automatically. The following verifies the first 1000 samples from the NuminaMath-LEAN dataset \citep{kimina-prover}.

\begin{lstlisting}[language=Python]
from datasets import load_dataset
from kimina_client import KiminaClient

client = KiminaClient()

client.run_benchmark(dataset_name="AI-MO/NuminaMath-LEAN", 
                     n=1000,
                     batch_size=8,
                     max_workers=10)
\end{lstlisting}

\newpage

For a more controlled and flexible approach, the following example shows how to manually prepare the data and call the core \texttt{check} function directly:
\begin{lstlisting}[language=Python]
from datasets import load_dataset
from kimina_client import KiminaClient
from kimina_client.models import Snippet

# Load the dataset
dataset = load_dataset("AI-MO/NuminaMath-LEAN", split="train")
dataset = dataset.filter(lambda x: x["ground_truth_type"] == "complete")
dataset = dataset.select(range(1000))

# Deduplicate on uuid
seen = set()
def keep_first(ex):
    u = ex["uuid"]
    if u in seen:
        return False
    seen.add(u)
    return True

dataset = dataset.filter(keep_first)

# Send to the Kimina Lean Server for validation
client = KiminaClient()

snippets = [
    Snippet(id=str(dataset[i]["uuid"]), code=dataset[i]["formal_ground_truth"])
    for i in range(len(dataset))
]

resp = client.check(snippets, timeout=120)

# Check the validation results for sample with a specific uuid
sample_uuid = "84f26e70-3dfd-589b-b7d0-7792576f0cc9"
for r in resp.results:
    if r.id == sample_uuid:
        print(f"Sample {sample_uuid} status: {r.analyze().status.value}")
        break
\end{lstlisting}

\newpage

\subsection{Proof Data Extraction}
\label{appendix:data-extraction}
This example demonstrates the client package's capability to process Lean's infotree output into a structured sequence of proof steps, each containing a tactic and its corresponding tactic states.

As an example, we take the following proof from the NuminaMath-LEAN dataset \citep{kimina-prover}.
\begin{lstlisting}
import Mathlib

/- Given that the product \( a \cdot b \cdot c = 1 \), what is the value of the following expression?
$$
\frac{a}{a b + a + 1} + \frac{b}{b c + b + 1} + \frac{c}{c a + c + 1}
$$-/
theorem algebra_4013 {a b c : ℝ} (h : a * b * c = 1) (haux : 1 + a + a * b ≠ 0) :
    a / (a * b + a + 1) + b / (b * c + b + 1) + c / (c * a + c + 1) = 1 := by
  -- need ne_zero condition to perform division
  have : a * b * c ≠ 0 := by rw [h]; norm_num
  have ha : a ≠ 0 := left_ne_zero_of_mul <| left_ne_zero_of_mul this
  have hb : b ≠ 0 := right_ne_zero_of_mul <| left_ne_zero_of_mul this
  --  Multiply the second fraction by \(a\).
  conv => lhs; lhs; rhs; rw [← mul_div_mul_left _ _ ha]
  --  Multiply the third fraction by \(ab\).
  conv => lhs; rhs; rw [← mul_div_mul_left _ _ (mul_ne_zero ha hb)]
  -- Thus, we get:
  --  \[
  --  \frac{a}{ab + a + 1} + \frac{ab}{abc + ab + a} + \frac{abc}{abca + abc + ab}
  --  \]
  rw [show a * (b * c + b + 1) = a*b*c + a*b + a by ring]
  rw [show a*b*(c * a + c + 1) = a*b*c*a + a*b*c + a*b by ring]
  -- **Simplify the expression using \(abc = 1\):**
  rw [h, one_mul]
  ring_nf
  -- **Combine the terms with the same denominator:**
  rw [← add_mul]
  nth_rw 2 [← one_mul (1 + a + a * b)⁻¹]
  rw [← add_mul, show a * b + a + 1 = 1 + a + a * b by ring]
  exact mul_inv_cancel₀ haux
\end{lstlisting}

First, we use the Python client to call the server with the \texttt{infotree="tactics"} option to retrieve the raw proof trace. Then, we use helper functions provided by the client library to parse this trace into a clean list of proof steps.
\begin{lstlisting}[language=Python]
dataset = load_dataset("AI-MO/NuminaMath-LEAN", split="train")

sample = dataset[0]

client = KiminaClient()

snippets = [
    Snippet(id=str(dataset[0]["uuid"]), code=dataset[0]["formal_ground_truth"])
]

resp = client.check(snippets, timeout=120, infotree="tactics")

infotree = resp.results[0].response["infotree"]
header, body = split_snippet(sample["formal_ground_truth"])
intervals = extract_data(infotree, body)
\end{lstlisting}

The resulting \texttt{intervals} object is a list where each element represents a distinct proof step with three key fields: \texttt{goalsBefore}, \texttt{tactic} and \texttt{goalsAfter}. 

\newpage

The first five steps for the example proof are shown below:

\begin{lstlisting}
INTERVAL 1
-----
Goals before:
a b c : ℝ
h : a * b * c = 1
haux : 1 + a + a * b ≠ 0
⊢ a / (a * b + a + 1) + b / (b * c + b + 1) + c / (c * a + c + 1) = 1
-----
Tactic:
by
  -- need ne_zero condition to perform division
  have : a * b * c ≠ 0 :=
-----
Goals after:
a b c : ℝ
h : a * b * c = 1
haux : 1 + a + a * b ≠ 0
⊢ a * b * c ≠ 0
--------------------
INTERVAL 2
-----
Goals before:
a b c : ℝ
h : a * b * c = 1
haux : 1 + a + a * b ≠ 0
⊢ a * b * c ≠ 0
-----
Tactic:
 by rw [h];
-----
Goals after:
a b c : ℝ
h : a * b * c = 1
haux : 1 + a + a * b ≠ 0
⊢ 1 ≠ 0
--------------------
INTERVAL 3
-----
Goals before:
a b c : ℝ
h : a * b * c = 1
haux : 1 + a + a * b ≠ 0
⊢ 1 ≠ 0
-----
Tactic:
 norm_num
-----
Goals after:
a b c : ℝ
h : a * b * c = 1
haux : 1 + a + a * b ≠ 0
this : a * b * c ≠ 0
⊢ a / (a * b + a + 1) + b / (b * c + b + 1) + c / (c * a + c + 1) = 1
--------------------
INTERVAL 4
-----
Goals before:
a b c : ℝ
h : a * b * c = 1
haux : 1 + a + a * b ≠ 0
this : a * b * c ≠ 0
⊢ a / (a * b + a + 1) + b / (b * c + b + 1) + c / (c * a + c + 1) = 1
-----
Tactic:

  have ha : a ≠ 0 := left_ne_zero_of_mul <| left_ne_zero_of_mul this
-----
Goals after:
a b c : ℝ
h : a * b * c = 1
haux : 1 + a + a * b ≠ 0
this : a * b * c ≠ 0
ha : a ≠ 0
⊢ a / (a * b + a + 1) + b / (b * c + b + 1) + c / (c * a + c + 1) = 1
--------------------
INTERVAL 5
-----
Goals before:
a b c : ℝ
h : a * b * c = 1
haux : 1 + a + a * b ≠ 0
this : a * b * c ≠ 0
ha : a ≠ 0
⊢ a / (a * b + a + 1) + b / (b * c + b + 1) + c / (c * a + c + 1) = 1
-----
Tactic:

  have hb : b ≠ 0 := right_ne_zero_of_mul <| left_ne_zero_of_mul this
-----
Goals after:
a b c : ℝ
h : a * b * c = 1
haux : 1 + a + a * b ≠ 0
this : a * b * c ≠ 0
ha : a ≠ 0
hb : b ≠ 0
⊢ a / (a * b + a + 1) + b / (b * c + b + 1) + c / (c * a + c + 1) = 1
\end{lstlisting}

\end{document}